%% file: raptor.tex
\documentclass[a4paper]{article}
\usepackage{amsmath, amsfonts}
\usepackage{algorithmicx}
\usepackage{multirow}
\usepackage{subcaption}
\usepackage{tikz}
\usetikzlibrary{
	graphs,
	automata,
	positioning,
	calc,
	datavisualization,
	datavisualization.formats.functions,
	quotes,
	shapes.geometric,
	shapes.symbols,
	patterns,
	decorations.pathreplacing,
    external
}
\usepackage{booktabs}
\usepackage{xurl}
\usepackage{graphicx}
\usepackage{hyperref}

\title{Raptor: Distributed Scheduling for Serverless Functions}
\author{
		Kevin Exton \\
		\href{mailto:kevin.exton@student.unimelb.edu.au}{kevin.exton@student.unimelb.edu.au}
		\and
		Maria A. Rodriguez \\
		\href{mailto:marodriguez@unimelb.edu.au}{marodriguez@unimelb.edu.au}}
\date{\today}
\begin{document}
\maketitle
\begin{abstract}
	To support parallelizable serverless workflows in applications like media processing, we have prototyped a distributed scheduler %
	called Raptor that reduces both the end-to-end delay time and failure rate of parallelizable serverless workflows. As modern %
	serverless frameworks are typically deployed to extremely large scale distributed computing environments by major cloud providers, %
	Raptor is specifically designed to exploit the property of statistically independent function execution that tends to emerge at %
	very large scales. To demonstrate the effect of horizontal scale on function execution, our evaluation demonstrates %
	that mean delay time improvements provided by Raptor for RSA public-private key pair generation can be accurately %
	predicted by mutually independent exponential random variables, but only once the serverless framework is deployed in %
	a highly available configuration and horizontally scaled across three availability zones.
\end{abstract}
\section{Introduction}
Serverless computing is a natural extension of the resource-as-a-service model that defines modern cloud computing trends. %
These platforms promise ad-hoc application deployments and automatic horizontal scaling that dynamically matches user application %
resource provisioning to instantaneous demand. This offers cloud providers with the flexibility to multiplex many users on the %
same underlying infrastructure, and it frees developers of the responsibility for dimensioning and provisioning %
compute resources for their applications. Application scheduling is key to satisfy the requirements of both serverless users and providers. Low-quality scheduling discourages developers from adopting serverless computing as it %
can severely impact the latency and throughput characteristics of an application, it can also make serverless computing %
less appealing for providers if compute resources can not be utilized as efficiently as traditional cloud-based virtual machines, or %
container deployments. 

Idiomatically, serverless computing services are exposed to users via API endpoints %
that abstract away the request routing and resource provisioning operations that must be performed by %
the microservices that make up the control plane of a serverless computing platform. These abstractions allow providers to %
deploy serverless computing services on the same shared compute infrastructure that is used to %
power other cloud offerings. This compute infrastructure is typically enormous in scale with individual %
sites taking up enough space that they are referred to as \emph{warehouse scale computer}~\cite{barroso_datacenter_2019}.

It is well established in the literature that job scheduling for systems at this scale has unique and unexpected performance %
characteristics. Chief among them the \emph{straggling effect}, which skews the end-to-end delay or response time %
distribution of scheduled jobs to have heavy tails~\cite{dean_tail_2013}. As an example, the average execution time of serverless functions from a two-week representative trace of Azure Functions invocations~\cite{shahrad_serverless_2020} exhibits a heavy-tailed distribution with a squared coefficient of variation of approximately %
11.4 and a mean of approximately 0.7 seconds. Although much of this variation is attributed to function cold-starts, many other contributing %
factors have been identified in the large-scale environments that cloud providers operate at~\cite{dean_tail_2013}.

To treat straggling for parallelizable serverless workflows in applications like video processing~\cite{fouladi_encoding_2017}, linear algebra~\cite{shankar_serverless_2020}, and %
machine learning~\cite{wang_serverlessml_2019}, we propose and prototype a distributed speculative-execution based scheduler\footnote{Source code available %
at~\cite{stochastic_scheduler_2024}.} called Raptor. Speculative execution has a strong track record as an effective straggler mitigation technique for data processing %
workloads~\cite{anan_effective_2013, ren_hopper_2015} on traditional frameworks like Spark~\cite{zaharia_resilient_2012} and Dryad~\cite{isard_dryad_2007}. Our design builds on the %
design and architecture of other performance-oriented serverless computing frameworks like Wukong~\cite{carver_wukong_2020}, SAND~\cite{akkus_sand_2018}, and %
Nightcore~\cite{jia_nightcore_2021}. These frameworks all eschew traditional container-isolation for lighter-weight thread or process-based isolation so that many %
different functions can be multiplexed on each container managed by the serverless computing framework. Raptor emphasizes the role that traditional operating system %
job controls can play in serverless computing frameworks that use process-based isolation for function invocations. In particular, Raptor uses POSIX job control %
signals to implement invocation \emph{preemption} so that function execution can be more efficiently time-shared across many parallel serverless worker nodes. %
Raptor also implements a \emph{fork} primitive that serves a similar role to the \texttt{fork()} system call. Using this primitive, invocations of parallelizable workflows %
and idempotent functions can \emph{optionally} be configured to replicate themselves across the distributed computing environment. Analogously to \texttt{fork()}, %
replicas can be connected together so that they can communicate with each other over the network. We use this communication channel between replicas to implement a %
state-sharing stream so that the state of the serverless function or workflow can be synchronized across all function replicas in the network. Our prototype has been designed to %
be deployed as a stand-alone container that can extend the feature set of traditional open-source serverless frameworks like %
OpenWhisk, Knative, or OpenFaaS with lighter weight process-based function isolation. 

We provide additional relevant background on job scheduling for distributed data processing and the design of modern performance-oriented serverless computing frameworks in %
section~\ref{s:background}. We give a detailed description of the design and key features of Raptor in section~\ref{s:ch5-Raptor}. We evaluate our prototype in %
section~\ref{s:ch5-evaluation}, before giving our conclusions in section~\ref{s:conclusions}. Our evaluations of Raptor were performed on the OpenWhisk serverless computing %
framework deployed across three Google Compute Platform (GCP) availability zones in a highly-available configuration. Our results demonstrate that Raptor %
simultaneously increases the reliability and reduces the end-to-end response times of serverless workflows consisting of tasks with a sufficient degree of randomness. %
To emphasize the effect of randomness on parallelizable workflow response times, and to demonstrate the role that horizontal scale plays in increasing %
the variance of workflow response times, we evaluate Raptor under a load of parallelizable RSA key generation requests serviced by the open source command-line %
utility \texttt{ssh-keygen} in section~\ref{ss:keygeneration}. The experiments in this section demonstrate that even in our small environment of no more than %
15 serverless worker nodes, increasing the scale of the system from 5 workers deployed in one availability zone to 15 workers deployed over three availability %
zones dramatically increases the variance and unpredictability of generating RSA keys in parallel. At 5 worker nodes deployed in just one availability zone, the generation of %
RSA keys in parallel has relatively predictable response times that almost completely eliminate the performance benefits of Raptor. At 15 nodes deployed over %
three availability zones, the generation of RSA keys in parallel is unpredictable enough that Raptor provides a reduction in invocation response times that is almost %
identical to what is predicted by independent exponentially distributed key generation times (or completely random execution times). We argue that the increase %
in variance and unpredictability associated with this increase in horizontal scale demonstrated by our experiments suggests that Raptor will %
perform better as system \emph{scale-out} in size, and that our design has properties that make it a good candidate for improving application %
performance on very large-scale serverless framework deployments such as those provided by major cloud-computing vendors.

\section{Background and Related Work}\label{s:background}
Serverless frameworks that have been developed with performance as a key consideration often forgo the traditional container isolation %
model of serverless computing for an operating system process or thread-based isolation model~\cite{akkus_sand_2018,jia_nightcore_2021,kotni_faastlane_2021,carver_wukong_2020}. %
Under these frameworks, a dedicated function executor is started inside of every container or VM that is spawned by the resource provisioning framework (e.g., %
Kubernetes~\cite{Kubernetes23}, YARN~\cite{vavilapalli_apache_2013}, Mesos~\cite{hindman_mesos_2011}, etc.). Each of these dedicated function executors is then %
initialized with the necessary instruction data to execute invoked user functions, usually, by pulling invocations from a collection of function invocation %
queues~\cite{jia_nightcore_2021, ow_scheduler_2020}. These dedicated executor daemons allow serverless computing frameworks to pool more than one user defined %
serverless function in each container and provides serverless computing frameworks with additional compute resource multiplexing capabilities via fine-grained %
operating system provided job controls. SAND~\cite{akkus_sand_2018} leverages the distributed nature of these dedicated function %
executors to short-circuit the data pathway of serverless workflows composed of a chain of serverless functions. Wukong~\cite{carver_wukong_2020} takes advantage %
of the tight integration between these distributed function executors with the core function scheduling components to more efficiently schedule serverless workflows %
that can be represented by a directed acyclic graph (DAG) of serverless functions.

This distributed architecture of function executors bears a strong similarity to the architecture of traditional data processing frameworks like %
MapReduce~\cite{dean_mapreduce_2008}, Spark~\cite{zaharia_resilient_2012}, and Dryad~\cite{isard_dryad_2007}. For parallelizable serverless workflows, especially, %
data processing workflows such as video processing~\cite{fouladi_encoding_2017}, linear algebra~\cite{shankar_serverless_2020}, and machine learning~\cite{wang_serverlessml_2019}, %
this implies that the straggling phenomenon that is known to plague distributed data processing frameworks at scale~\cite{gill_tails_2020} will also affect %
distributed serverless function execution at scale. In fact, the heavily skewed distribution of function execution times present in the 2019~\cite{shahrad_serverless_2020} and %
2021~\cite{zhang_faster-serverless_2021} Azure Functions trace supports this hypothesis. The 2019 Azure trace exhibits serverless function response times with a mean of %
approximately 0.7 seconds and a squared coefficient of variation of around 11.7, while the 2021 Azure trace exhibits mean response times of around 3 seconds and %
with a squared coefficient of variation of around 30.6. Larger squared coefficients of variation indicate more variation in function response times with respect to the mean %
and the increase between 2019 and 2021 indicates that the distribution tail of function response times in the Azure Functions environment has grown longer in this interval %
and that the straggling problem has become worse.

Speculative execution is a common technique used by job schedulers to mitigate straggling for traditional distributed data %
processing~\cite{lei_crest_2011,phan_energy-driven_2017,anan_effective_2013,ren_hopper_2015, dean_mapreduce_2008} and has a strong track record of reducing %
the average response times of parallelizable data processing jobs. The main cost of speculative execution in distributed environments is the excess %
resources consumed from unnecessarily scheduled copies. This excess resource consumption can reduce both the aggregate capacity of the system as a whole, %
and increase energy consumption. A wide variety of strategies have been employed in the literature in an attempt to limit excess resources %
consumed by speculative execution including: delayed speculation~\cite{zaharia_improving_2008,ananthanarayanan_reining_2010}, %
straggler prediction~\cite{dai_improvedstraggler_2017,farhat_stochastic_2018}, and speculation budgeting~\cite{ren_hopper_2015,anan_effective_2013}. %
Perhaps the simplest technique for mitigating excess resource consumption, however, is \emph{preemption}. In fact, %
under Poisson arrival and mutually independent exponential service time assumptions, preemption is sufficient to %
completely eliminate the reduction in system capacity that would otherwise be associated with speculative execution~\cite{gardner_reducing_2015}.

We have designed Raptor to mitigate the tails of the response time distribution of parallelizable serverless workflows using speculative execution. Unlike %
other speculative execution approaches that are used by distributed data processing frameworks that we have outlined here, Raptor is designed to %
mitigate the cost of speculatively scheduled copies by means of \emph{preemption}. As Raptor is designed as a distributed function executor, function %
invocation preemption can be implemented using traditional operating system job controls, and our prototype implements preemption by %
isolating each function invocation into its own process group and managing each invocation with POSIX job control signals. As the total amount of %
additional resources consumed by speculative execution will be sensitive to preemption signal latency, Raptor is designed to execute parallelizable serverless workflows in %
distributed execution groups called ``flights''. The state of a serverless invocation is shared between all instances of Raptor in a flight in a peer-to-peer %
fashion so that resources can be freed immediately after at least one member in the flight finishes all of the tasks assigned to the serverless invocation. %
Internally, Raptor represents the dependencies between functions in an invoked serverless workflow using a directed acyclic graph, this enables Raptor %
to short-circuit the data pathway of functions that are chained together in a fashion that is similar to SAND~\cite{akkus_sand_2018}, with the added %
flexibility of chaining functions that produce a single output into functions that require multiple inputs using an approach that is similar to Wukong~\cite{carver_wukong_2020}. %
Unlike the strategies employed by schedulers in other serverless frameworks, Raptor will speculatively execute all tasks that have been assigned to a flight %
over all flight members. This process ensures that the entire flight can be preempted as soon as at least one of the flight members completes the last task in %
its assigned serverless workflow.

\section{The Scheduler -- ``Raptor''}\label{s:ch5-Raptor}
Raptor is designed to extend, in a backwards-compatible way, the routing and resource management infrastructure of existing serverless frameworks. %
Since the integration point for OpenWhisk is easy to access and well documented~\cite{ow-action-interface}, we have integrated our prototype of Raptor with OpenWhisk's routing %
and queuing services for our evaluation. In contrast, the equivalent integration point in Knative, called the Queue-Proxy~\cite{knative_architecture}, is an internal component %
to the Knative serving infrastructure, has limited documentation, and does not appear to have easily accessible points for extensions or plugins. Similarly, the %
OpenFaaS watchdog~\cite{openfaas_of_watchdog} is also an internal component of the OpenFaaS infrastructure with limited documentation and no %
easily accessible points for extensions or plugins.

\subsection{Function Invocation, Routing, and Scheduling in OpenWhisk}
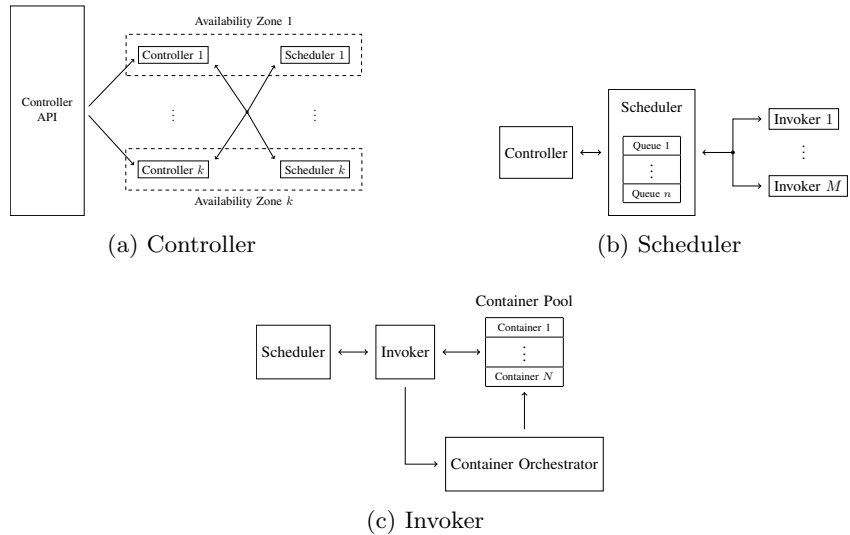
\begin{figure}
	\centering
	\subfloat[Controller\label{f:ch5-ow-controller}]{
		\resizebox{0.4\textwidth}{!}{
			\tikzsetnextfilename{ch5-controllerarch}
			\input{figures/controller-arch.tex}
		}
	}\hspace{0.1\textwidth}
	\subfloat[Scheduler\label{f:ch5-ow-scheduler}]{
		\resizebox{0.4\textwidth}{!}{
			\tikzsetnextfilename{ch5-schedulerarch}
			\input{figures/schedulerarch.tex}
		}
	}

    \vspace{1em}
	\subfloat[Invoker\label{f:ch5-ow-invoker}]{
		\resizebox{0.4\textwidth}{!}{
			\tikzsetnextfilename{ch5-invokerpool}
			\input{figures/InvokerPool.tex}
		}
	}
	\caption{OpenWhisk Architecture\label{f:ch5-ow-arch}}
\end{figure}

The OpenWhisk technology stack has three layers. At the bottom, there is a data persistence layer that, by %
default, is provided by CouchDB~\cite{couchdb}. In the middle, there is a messaging layer provided by Kafka~\cite{Kafka24}. The microservices that make up the core components of %
OpenWhisk are at the top of the stack: controller (Figure~\ref{f:ch5-ow-controller}), scheduler (Figure~\ref{f:ch5-ow-scheduler}), and invoker (Figure~\ref{f:ch5-ow-invoker}). %
Each of OpenWhisk's core microservices is responsible for managing a different aspect of the flow of data through the framework, in increasing %
order of complexity: the controller implements the OpenWhisk API, the invoker integrates with a container orchestrator %
to maintain a pool of compute resources, and the scheduler uses queues to manage the cold-start frequency of functions (Figure~\ref{f:ch5-ow-arch}). %
The controller and scheduler microservices are designed to be co-located and deployed in a configuration with up to $k$ replicas that load share the %
API traffic. OpenWhisk invokers are, instead, designed to be co-located with the hardware resources where containers are deployed but can %
also scale horizontally so that arriving requests can be load-balanced across $M$ replicas. Communication between each of the core microservices is %
mostly performed via Kafka messages, except that, gRPC~\cite{grpc_2024} is used between invokers and schedulers to pull user function invocation %
requests from the head of a queue in the scheduler~\cite{ow_scheduler_2020}. Akka~\cite{akka24} is used to provide the lower level %
networking primitives that connect the OpenWhisk technology stack together.

To avoid confusion, we must mention that there are two different versions of the OpenWhisk architecture. The %
original architecture, released to the community prior to 2016, and the new (or 2.0) architecture introduced by the merging of a new %
scheduler component in November of 2020~\cite{openwhisk_scheduler_2020}, although the official 2.0 release of the OpenWhisk framework was %
delayed until April of 2024~\cite{ow2-release}. Our prototype, Raptor, has been designed to extend the newer OpenWhisk 2.0 architecture.

\subsection{Raptor Architecture}
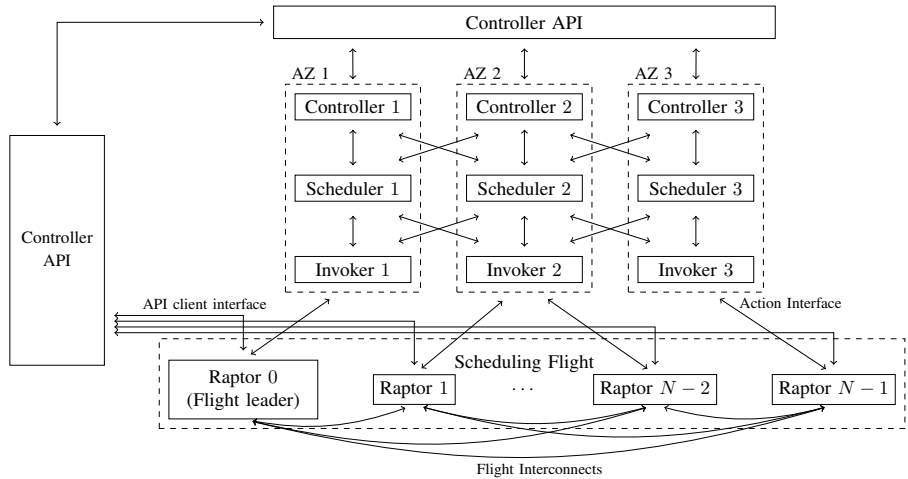
\begin{figure}
	\centering
	\resizebox{\textwidth}{!}{
		\tikzsetnextfilename{RaptorArchitecture}
		\input{figures/Raptor_Architecture.tex}
	}
	\caption{Distributed Scheduling Architecture\label{f:ch5-RaptorArchitecture}}
\end{figure}

Figure~\ref{f:ch5-RaptorArchitecture} depicts the end-to-end architecture of a horizontally scaled Raptor \emph{flight} including the %
integration points with the core OpenWhisk services. A flight has $N$ parallel instances of Raptor connected together %
in a peer-to-peer mesh. Flights are dynamically scheduled via an action \emph{fork} when a user invokes a serverless function or workflow that %
has been configured for replicated execution. The first instance of Raptor that is assigned such an action is %
called the \emph{flight leader} and has an index of $0$. The flight leader is responsible for \emph{forking} the function by recursively %
invoking the action along with the user-provided parameters and some additional metadata so that the replicas can %
connect to each other and establish the state-sharing stream. Replicas are assigned %
an index greater than 0 by the flight leader as part of the action \emph{fork} procedure, the index of each instance within a %
flight is fixed for the execution time of the invocation but is dynamic across invocations. This means that %
each invocation of a serverless function or workflow can be assigned to a different group of %
parallel instances where the groups may overlap with each other, e.g., a sequence of identical invocations %
configured to run on a flight of size $2$ and scheduled to run on a cluster with %
$3$ containers has $\binom{3}{2}$ possible flight combinations. To keep the communication latencies between invocation replicas low, our state-sharing stream is %
implemented using the SCTP transport protocol. This allows us to establish ongoing SCTP associations \emph{once} on a function cold-start and multiplex all subsequent %
state-sharing streams over the SCTP streams provided by the transport protocol. As SCTP eliminates transport layer head-of-line blocking, %
all events broadcast on the state-sharing stream after a function cold-start are guaranteed to be delivered in one half network round-trip time. This ensures that %
function invocation data in a serverless workflow always takes the shortest path to its destination, and it ensures that all of the parallel instances of Raptor %
within a flight free their resource allocations immediately after an invocation has finished execution.

\begin{figure}
	\centering
    \tikzsetnextfilename{raptorInterfaces}
    \input{figures/raptorInterfaces.tex}
	\caption{Raptor-OpenWhisk Interfaces\label{f:ch5-RaptorInterfaces}}
\end{figure}
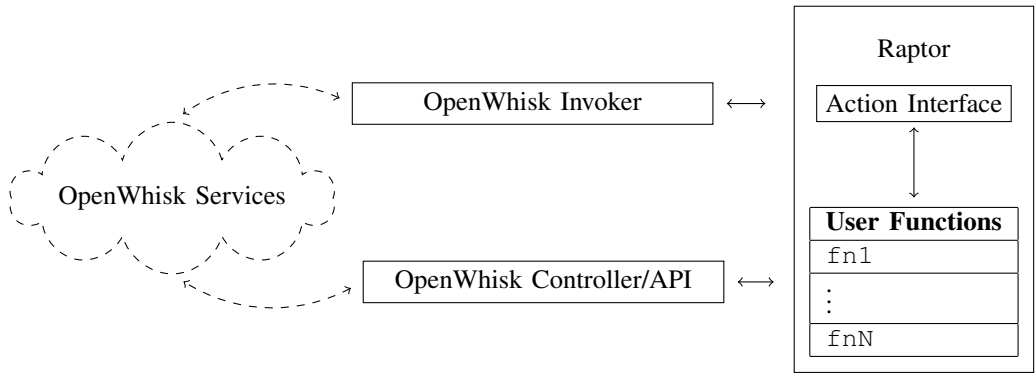

In our prototype, each instance of Raptor implements two interfaces towards the core OpenWhisk services: an action proxy server that %
receives activation requests from invokers over the OpenWhisk action interface, and an API client that sends requests to an OpenWhisk controller %
to implement the action \emph{fork} procedure (Figure~\ref{f:ch5-RaptorInterfaces}). %
Internally, each instance of Raptor implements an interface that language runtimes can integrate with. %
Raptor's runtime interface requires that each function invocation is isolated into its own process %
that inherits a set of isolated file descriptors provided by Raptor for IPC. This enables traditional operating %
system controls to be used to secure user data and ensures that our runtime interface is no less secure than %
the traditional container isolation model for serverless functions. Currently, our prototype only supports executing %
functions via system calls to \texttt{fork()} and \texttt{exec()}, which is sufficient for languages with short runtime %
initialization times like NodeJS and Lua. For languages like Python 3 with long runtime initialization times, we leave open the future option %
of implementing a fork-server action interface for function execution.

\subsection{Scheduling Procedure for Raptor}
The end-to-end function scheduling flow in a serverless framework with Raptor can be broken down %
into four phases:
\subsubsection{Build and Place an Action Manifest}
Raptor is designed to schedule groups of user-defined functions within a single container, either in series or in parallel, allowing it to take full %
advantage of the forking mechanism for serverless function or workflow invocations. As such, invocations arriving at an instance of Raptor should %
provide a manifest of user-defined serverless functions called an \emph{action manifest} that indexes the functions to be executed by where they are located, defines any %
dependencies they may have between each other, and specifies the degree of concurrency or number of replicas that this invocation should be executed with. %
Table~\ref{t:ch5-action-manifest} is an example of an action manifest containing four functions with dependencies such that fn1 must complete before fn2 and fn3 can begin, and %
fn2 and fn3 must complete before fn4 can begin; as fn2 and fn3 do not depend on each other, they are allowed to %
execute in parallel. As this manifest has a concurrency of 2, the flight leader for every invocation %
of this manifest will replicate itself exactly once so that there will be two replicas of the manifest executing %
simultaneously on the serverless computing cluster. The action manifest and all of the function code needed to execute an action manifest are bundled together %
and delivered to the executor once upon a cold start. For our prototype on OpenWhisk, this is implemented by supplying a %
zipped archive of files via an HTTP route in the action interface and is the same mechanism as that for packaging and deploying functions with third-party dependencies on %
serverless frameworks.
\begin{table}
	\centering
	\caption{Action Manifest of Four Functions}
	\label{t:ch5-action-manifest}
    \begin{tabular}{cccc}\toprule
        \textbf{Name} 		& \textbf{Location} 	& \textbf{Dependencies} 				& \textbf{Concurrency}\\
        \midrule
        \texttt{fn1:main} & \texttt{<PATH>} 		& \texttt{[]} 							& 2\\
        \texttt{fn2:main} & \texttt{<PATH>} 		& \texttt{[fn1:main]} 					& 2 \\
        \texttt{fn3:main} & \texttt{<PATH>} 		& \texttt{[fn1:main]} 					& 2\\
        \texttt{fn4:main} & \texttt{<PATH>} 		& \texttt{[fn2:main,fn3:main]} 	& 2 \\
        \bottomrule
    \end{tabular}
\end{table}

\subsubsection{Fork The Manifest}
Once an action sandbox has been initialized with a manifest and its associated code, the executor within that sandbox %
is ready to receive invocation requests. If an invocation request arrives that identifies the receiving executor %
as having a flight index of 0 and the action manifest has a concurrency greater than one, then the %
flight leader will \emph{fork} the action invocation by wrapping the user-supplied invocation parameters %
in an execution context and recursively invoking the action via the serverless API with the supplied %
execution context. Each execution context is assigned a UUID so that repeated invocations of the same context %
can be distinguished from repeated invocations of different contexts, each execution context must %
provide an address where the leader can be reached at so that a state-sharing stream can be %
established between all of the executors in a flight, and each recursive invocation of an execution context must %
be assigned a different index by the leader that is greater than 0 so that followers can uniquely identify each other. %
Table~\ref{t:ch5-executionctx} provides an example of the metadata added to invocation requests to implement action \emph{forking}. The leader %
is responsible for ensuring that all of the following invocations are correctly joined into the distributed state-sharing stream. If %
the leader of a flight consisting of $N$ replicas fails after $M < N-1$ followers have joined the state-sharing stream, %
then the flight will operate at a reduced size of $M$ replicas and the remaining $N-M-1$ followers will fail gracefully.
\begin{table}
	\centering
	\caption{Execution Context\label{t:ch5-executionctx}}
    \begin{tabular}{cc}\toprule
        \textbf{Parameter Name} & \textbf{Parameter Value} \\
        \midrule
        Context UUID & \texttt{<UUID>} \\
        Leader Address & \texttt{<IP Address>} \\
        Follower Index & \texttt{<Integer greater than 0>} \\
        User Parameter 1 & \texttt{<Param 1>} \\
        $\vdots$ & \\
        User Parameter N & \texttt{<Param N>} \\
        \bottomrule
    \end{tabular}
\end{table}

\subsubsection{Execute the Manifest}
Each parallel executor is designed to execute functions from the action manifest one at a time. As such, an execution sequence %
must be constructed from the action manifest as provided. To do so, each executor will first construct %
a directed acyclic graph of functions from the dependency information provided in the action manifest. Then, starting at the %
end of the graph, each executor applies an in-order tree traversal algorithm in the reverse direction to search for the first function in the graph that has all of %
its data dependencies satisfied; this becomes the next function in the execution sequence. Figure~\ref{f:ch5-am-dag} gives the directed acyclic %
graph for the action manifest in Table~\ref{t:ch5-action-manifest}. To make the execution sequences between different executors as independent %
as possible, we use the follower index assigned to the local executor to apply a cyclic shift to the search order of nodes %
in the directed acyclic graph. Table~\ref{t:ch5-am-sequence} lists example execution sequences that would be produced by %
two parallel executors of the action manifest in Table~\ref{t:ch5-action-manifest}.
\begin{figure}
	\centering
    \tikzsetnextfilename{ch5-manifestdag}
    \input{figures/manifest-dag.tex}
	\caption{Directed Acyclic Graph of an Action Manifest.}
	\label{f:ch5-am-dag}
\end{figure}
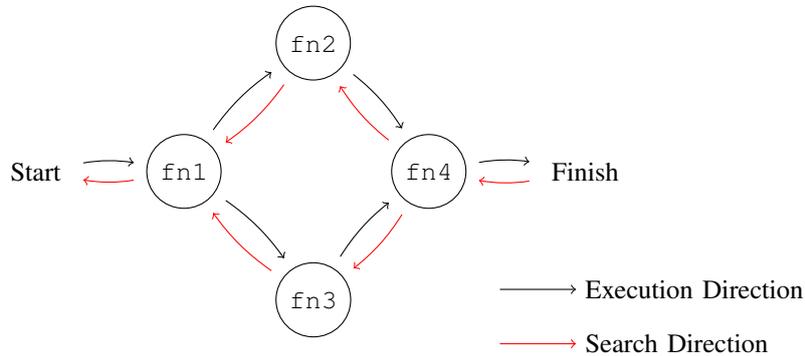
\begin{table}
	\centering
	\caption{Execution Sequence on Two Parallel Executors}
	\label{t:ch5-am-sequence}
    \begin{tabular}{rcccc} \toprule
                                                & 0 	& 1 	& 2 	& 3 \\
        \midrule
        \textbf{Executor 0}		& \texttt{fn1} & \texttt{fn2} & \texttt{fn3} & \texttt{fn4} \\
        \textbf{Executor 1} 	& \texttt{fn1} & \texttt{fn3} & \texttt{fn2} & \texttt{fn4} \\
        \bottomrule
    \end{tabular}
\end{table}

\subsubsection{Preempt The Manifest}
Each time one of the distributed execution instances completes one of the functions in the manifest, it will %
broadcast a notification containing the function output to all of the other members in the flight before %
proceeding to the next function in the sequence. If an executor receives a function output prior to the next function in the %
sequence completing locally, it will stop the function from further execution. If the function %
is not yet in a started state, it will not be scheduled to start in the future. If the function has already started, %
then job control signals will be sent to the subprocess group associated with it to gracefully reclaim all of the %
the system resources that it has been assigned. If the function has already completed by the %
time a function output arrives on the network, then two parallel executors have completed a function simultaneously %
and the local executor will update its state with data from the first event that does not contain an error, %
if no errors have occurred then it simply discards the event. 

\section{Performance Evaluation}\label{s:ch5-evaluation}

\subsection{OpenWhisk Performance Overhead}
As the performance of Raptor is sensitive to the delay overhead added by the OpenWhisk core microservices, our evaluation begins by first tuning %
these microservices to ensure predictable performance at scale. The workflows we evaluated were then chosen because they have %
mean execution times at least one order of magnitude larger than this overhead. This ensures that the contribution made by %
the OpenWhisk core microservices to the end-to-end delay time of the workflow is negligible.

\subsubsection{OpenWhisk Configuration}
A high-performing deployment of OpenWhisk must ensure that all of the major components in the technology stack are tuned to work well %
both individually and together. A complete performance tuning procedure for OpenWhisk requires us to methodically work our way through %
the technology stack, tuning each component one at a time with careful consideration for how they communicate with each other and %
how resources must be shared between them. We give an overview of our OpenWhisk configuration %
starting from the hardware deployed in our cloud environment and working our way up the OpenWhisk technology stack.

Our currently preferred cloud environment is the Google Cloud Platform (GCP) Compute Engine. Our %
cluster configuration consists of a total of 21 virtual machines (VMs), with 3 VMs dedicated to Kubernetes controllers, %
3 VMs dedicated to Kafka and the OpenWhisk controller and scheduler microservices, and 15 VMs designated as %
worker nodes that provide the compute resources for user functions. Our cluster is deployed in a high-availability (HA) configuration %
over 3 availability zones where in each availability zone we deploy: 1 VM for a Kubernetes controller, 1 VM for Kafka and the OpenWhisk %
controller and scheduler microservices, and 5 worker node VMs, for a total of 7 VMs in each availability zone. %
Table~\ref{t:ch5-ClusterConfiguration} enumerates the hardware deployed for each of these three node types.
\begin{table}
	\centering
	\caption{Cluster Configuration\label{t:ch5-ClusterConfiguration}}
	\resizebox{\textwidth}{!}{
		\begin{tabular}{rccccccc}\toprule
			& GCP VM Type & No. vCPUs & Avail. Memory & Avail. Disk & Disk Type & No. VMs & No. VMs per AZ\\
			\cmidrule{2-8}
			Kubernetes Controller 	& n2-standard-2 & 2 & 8GiB & 64GiB & SSD & 3 & 1\\
			OpenWhisk Controller and Scheduler & e2-highmem-2 & 2 & 16GiB & 32GiB & SSD & 3 & 1\\
			Worker Node& e2-small & 0.5 - 2 & 2GiB & 32GiB & HDD & 15 & 5\\
			\bottomrule
		\end{tabular}
	}
\end{table}

We use CouchDB as our database for the OpenWhisk persistence layer, which we deploy on our worker nodes rather than our %
OpenWhisk control nodes to prevent CouchDB and Kafka from cannibalizing each other's file system cache performance. This is because the performance %
of Kafka is highly sensitive to the file system cache~\cite{kafka_fs_2024}. To spread database reads and writes out over all 15 worker nodes, %
we configure CouchDB databases to use 5 shards that are each replicated 3 times. Where possible, we have made an %
effort to tune both Akka and Kafka at the messaging layer, and make adjustments to the Java VM to improve %
performance. Table~\ref{t:ch5-owconf} summarizes the major configuration adjustments we have made to the OpenWhisk services to provide a good %
balance between high availability and low latency.
\begin{table}
	\centering
	\caption{OpenWhisk Configuration}
	\label{t:ch5-owconf}
	\resizebox{\columnwidth}{!}{
		\begin{tabular}{ccccc}\toprule
			\textbf{Component} & \textbf{Scope} & \textbf{Option Name} & \textbf{Option Value} & \textbf{Default Value} \\
			\midrule
			\multirow{2}{*}{Kafka} & \multirow{2}{*}{All OpenWhisk Topics} & segment.bytes & 2147483647 & 536870912 \\
			&                      				   & retention.bytes & 4294967296 & 1073741824 \\
			\midrule
			\multirow{6}{*}{Akka} & \multirow{3}{*}{All OpenWhisk Microservices} & akka.io.tcp.batch-accept-limit & 32 & 10 \\
			&	& akka.scheduler.tick-duration & 1ms & 10ms \\
			&	& akka.scheduler.ticks-per-wheel & 4096 & 512 \\
			\cmidrule{2-5}
			& \multirow{3}{*}{All Thread Pool Executor Dispatchers} & core-pool-min-size & 1 & 2\\
			& & core-pool-size-factor & 1.0 & 2.0 \\
			& & core-pool-size-max & 16 & 32 \\
			\midrule
			\multirow{7}{*}{JVM Options} & \multirow{7}{*}{All OpenWhisk Microservices} & UseZGC & \checkmark & \\
			& & AlwaysPreTouch & \checkmark &  \\
			& & AggressiveHeap & \checkmark &  \\
			& & -server & \checkmark &  \\
			& & UseFastAccessorMethods & \checkmark &  \\
			& & UseStringCache & \checkmark &  \\
			& & OptimizeStringConcat & \checkmark &  \\
			\bottomrule
		\end{tabular}
	}
\end{table}

\subsubsection{OpenWhisk Performance Overhead Evaluation}
\begin{figure}
	\centering
	\subfloat[Low vs high-availability Control Plane Overhead\label{f:ch5-cpoverhead-lvh}]{
		\includegraphics[width=0.8\textwidth]{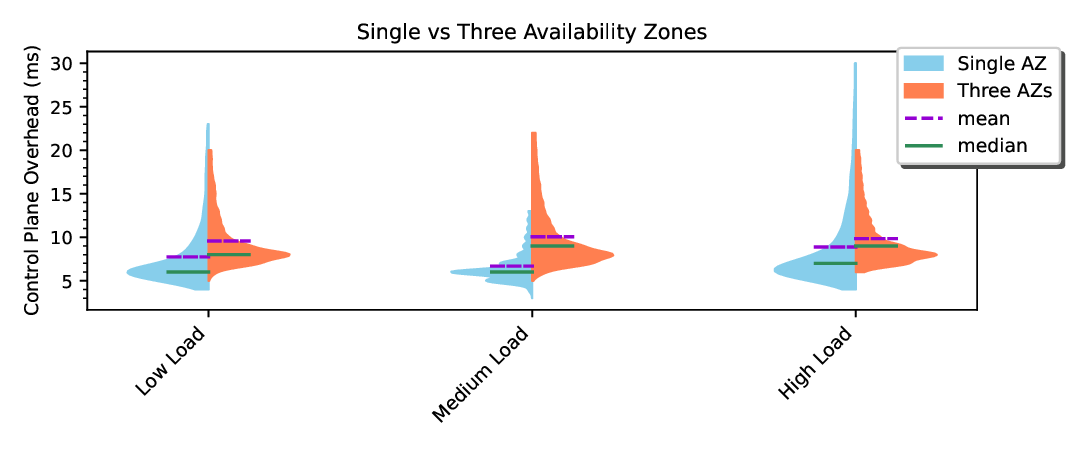}
	}
	
	\vspace{1em}
	\subfloat[high-availability Control Plane Overhead\label{f:ch5-cpoverhead-h}]{
		\includegraphics[width=0.35\textwidth]{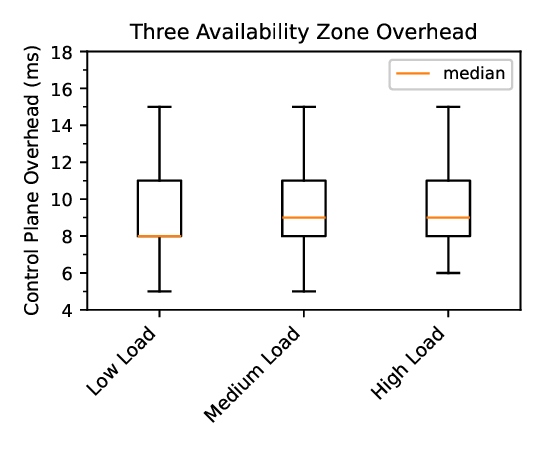}
	}\hspace{0.5in}
	\subfloat[Low Availability Control Plane Overhead\label{f:ch5-cpoverhead-l}]{
		\includegraphics[width=0.35\textwidth]{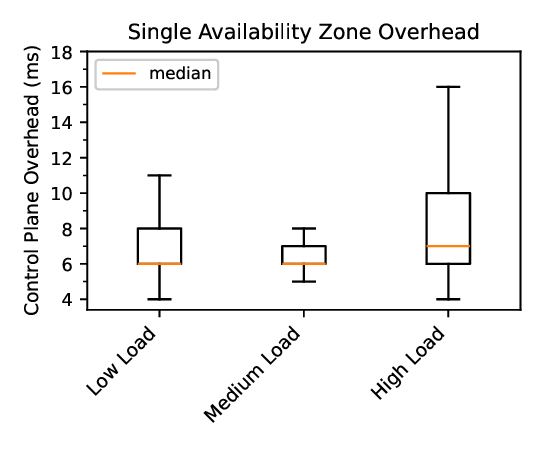}
	}
	\caption{OpenWhisk Overheads\label{f:ch5-cp-overheads}}
\end{figure}

Figure~\ref{f:ch5-cp-overheads} illustrates the OpenWhisk control plane overhead with our configuration and demonstrates that %
although the control plane has a median delay overhead of around 9ms, it is extremely stable under a %
wide range of loads. %
Of note is the fact that median overhead increases by approximately 2ms when we scale out from a low availability %
deployment in a single availability zone, to a high-availability deployment spread out over three availability zones. %
Table~\ref{t:ch5-ow-cp-comparison} compares the control plane overhead of OpenWhisk with that of Il\'{u}vatar~\cite{fuerst_iluvatar_2023}, %
a serverless framework that has been developed with the intention of being used for research. The results demonstrate that the %
overall control plane overhead of OpenWhisk can be quite acceptable when properly tuned, %
\begin{table}
	\centering
	\caption{Control Plane Latency Overhead Comparison}
	\label{t:ch5-ow-cp-comparison}
	\resizebox{0.6\columnwidth}{!}{
		\begin{tabular}{ccccc}\toprule
			& 																								& Three AZs & One AZ 	& Il\'{u}vutar \\
			\midrule
			\multirow{2}{*}{Low Load} 		& Median 						& 8 ms  		& 6 ms 		& 2 ms\\
			& 90'th percentile 	& 14 ms 		& 12 ms 	& 3 ms \\
			\cmidrule{2-5}
			\multirow{2}{*}{Medium Load} 	& Median 						& 9 ms 			& 6 ms 		& 2.5 ms\\
			& 90'th percentile 	& 16 ms 		& 9 ms 		& 3.5 ms \\
			\cmidrule{2-5}
			\multirow{2}{*}{High Load} 		& Median 						& 9 ms 			& 7 ms 		&  2.5 ms\\
			& 90'th percentile 	& 15 ms 		& 15 ms 	& 10 ms \\   
			\bottomrule
		\end{tabular}
	}
\end{table}%
especially when considering the fact that %
the OpenWhisk architecture provides message delivery guarantees via the use of Kafka topics and that replicated high-availability %
deployments of the core microservices will add a non-negligible amount of communication overhead between microservices that %
may be located in different availability zones.

\subsection{Raptor Performance Evaluation}
The distributed flight-based scheduling approach adopted by Raptor performs best on parallelizable workloads since it is inspired by the speculative execution techniques %
commonly used by distributed data processing frameworks~\cite{ren_hopper_2015,zaharia_improving_2008,ananthanarayanan_reining_2010}. Therefore, we have evaluated the performance %
benefits of Raptor under the three different parallelizable workflows in sections~\ref{ss:keygeneration} and~\ref{ss:other-workloads} and have summarized the results in %
Table~\ref{t:ch5-responsetimes}. We argue that restricting the scope of Raptor to these types of workflows is not a significant limitation of our design as parallelizable %
jobs are commonly implemented using serverless functions~\cite{fouladi_encoding_2017,carver_wukong_2020,shankar_serverless_2020}, and %
serverless frameworks often use separate scheduling modes for different types of workflows; for instance, %
OpenWhisk action sequences~\cite{ow-action-sequence} and conductor actions~\cite{ow-conductor-actions}, %
AWS Step Functions~\cite{aws-step-functions}, Google Cloud Workflows~\cite{google-cloud-workflows}, and Azure Logic Apps~\cite{azure-logic-apps}.
\begin{table}
	\centering
	\caption{Response Times for Evaluated Workflows.}
	\label{t:ch5-responsetimes}
    \begin{tabular}{cccc}\toprule
                                                                                                &										& w/o Raptor 	& w/ Raptor 
\\
        \midrule
        \multirow{3}{*}{SSH Key Generation} 				& Median 						& 939 ms				& 674 ms\\
                                                                                                & Mean 							& 1335 ms				& 864 ms\\
                                                                                                & 90'th Percentile 	& 2887 ms				& 1721 
ms\\
        \cmidrule{2-4}
        \multirow{3}{*}{Word Count} 								& Median 						& 4126 ms				& 1920 ms \\
                                                                                                & Mean 							&	4296 ms				& 1954 ms 
\\
                                                                                                & 90'th Percentile	&	3039 ms				& 1416 ms \\
        \cmidrule{2-4}					 																	
        \multirow{3}{*}{Image Thumbnail Generation} & Median 						& 1673 ms				& 1492 ms \\
                                                                                                & Mean 							& 1653 ms				& 1474 ms 
\\
                                                                                                & 90'th Percentile	& 2040 ms				& 1872 ms \\
        \bottomrule
    \end{tabular}
\end{table}

\subsubsection{SSH Key Generation Workflow}\label{ss:keygeneration}
As Raptor is designed to work best under the assumption that parallel executors will have \emph{mutually independent} function execution %
times, we have designed a synthetic workload consisting of SSH public-private key pair generation tasks using the command line utility %
\texttt{ssh-keygen}. As SSH key generation must wait for there to be a sufficient amount of system entropy to source the random numbers required %
to generate the cryptographic keys, the execution time of SSH key generation will be different between any two parallel compute %
nodes as they will have independent sources of cryptographic entropy. As Raptor is designed for parallelizable jobs, %
our synthetic workload generates SSH key pairs two at a time in parallel, Table~\ref{t:ch5-ssh-kg-am} lists %
the action manifest for our key generation jobs. As OpenWhisk does not provide a native action API for building %
parallelizable jobs, we implement a simple NodeJS fork-join job coordinator that makes two concurrent requests for SSH key generation %
to the OpenWhisk API for every arriving job request and returns a result containing both outputs, Algorithm~\ref{a:ch5-coord} %
provides an algorithmic description of our coordinator. Our experiments are performed by generating a Poisson arrival sequence of job %
invocation requests that are injected into OpenWhisk over a period of 30 minutes.
\begin{table}
	\centering
	\caption{Key Generation Action Manifest}
	\label{t:ch5-ssh-kg-am}
	\resizebox{0.6\columnwidth}{!}{
		\begin{tabular}{cccc}\toprule
			\textbf{Name} 			& \textbf{Location} & \textbf{Dependencies} & \textbf{Concurrency} \\
			\midrule
			\texttt{ssh-keygen} & \texttt{<PATH>} 	& \texttt{[]} 					& 2 \\
			\texttt{ssh-keygen} & \texttt{<PATH>} 	& \texttt{[]}						& 2 \\
			\bottomrule
		\end{tabular}
	}
\end{table}

\begin{figure}[!ht]
	\centering
	\subfloat[high-availability\label{f:ch5-sshkeys-ha}]{
		\includegraphics[width=0.9\textwidth]{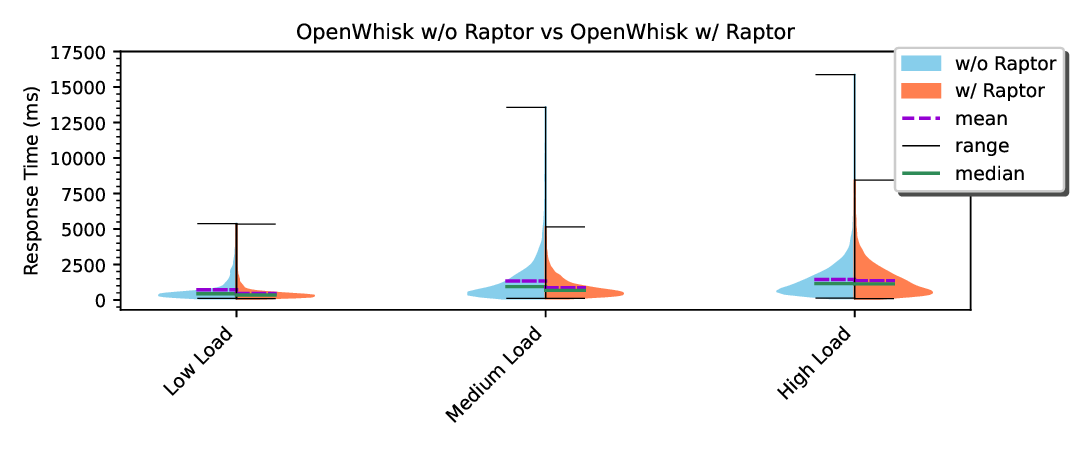}
	}
	
	\vspace{1em}
	
	\subfloat[Low Availability\label{f:ch5-sshkeys-la}]{
		\includegraphics[width=0.9\textwidth]{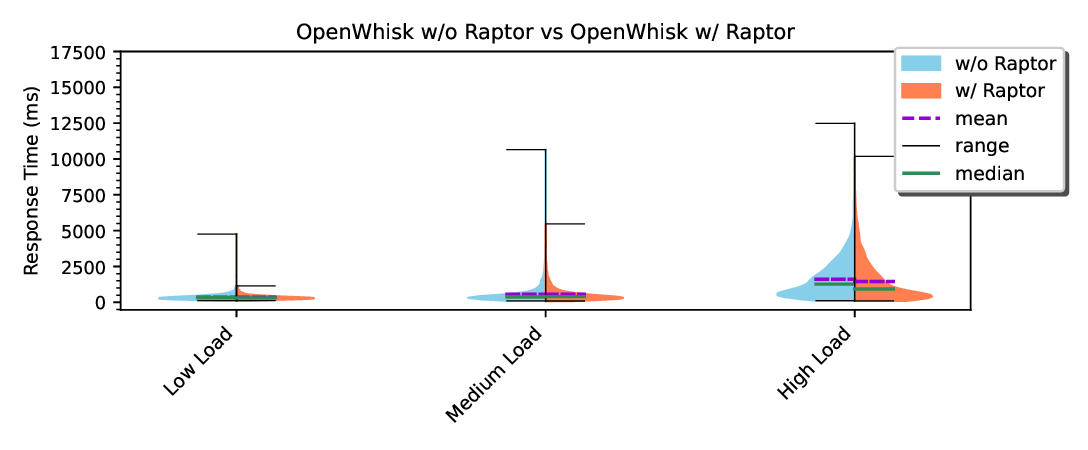}
	}
	\caption{Performance of OpenWhisk w/ Raptor vs Stock OpenWhisk}
	\label{f:ch5-sshkeys}
\end{figure}

Figure~\ref{f:ch5-sshkeys} illustrates the performance of SSH key generation on OpenWhisk both with and without Raptor, and for both %
low and high-availability deployments. %
As a general rule of thumb, Raptor on OpenWhisk performs best under moderate system loadings. This is because %
under a moderate system loading, the average waiting time in the Kafka message queue is still relatively low (on the same order of magnitude as %
the OpenWhisk control plane overhead), at the same time, there is a sufficient level of system load that the OpenWhisk schedulers (depicted %
in Figure~\ref{f:ch5-ow-scheduler}) will horizontally scale the number of parallel action containers to a high enough number that the %
parallel executors for each job invocation are likely to be placed on compute nodes that have \emph{mutually independent} sources of %
cryptographic entropy. At low loads, OpenWhisk will not scale out the number of concurrent action containers enough to ensure that %
each invocation will have independent function execution time characteristics, while at high loads, waiting times in the Kafka message %
queue will begin to dominate the job response time limiting the effectiveness of the replication strategy. Of special note, is the fact that under %
a low-availability configuration, the mean response time under Raptor is only 1\% shorter than the mean response time under stock OpenWhisk on %
a moderately loaded cluster, which implies that Raptor provides no performance benefits over stock OpenWhisk when there are only 5 worker nodes. %
However, under a high-availability deployment, Raptor reduces mean response times by approximately 35\% relative to stock OpenWhisk. This %
is almost exactly the performance improvement we would expect to see if job execution times on parallel nodes were %
\emph{mutually independent} and \emph{exponentially distributed} which can be found by the equation:
\begin{equation*}
	\begin{split}
	\frac{E\left[T_{\text{Raptor}}\right]}{E\left[T_{\text{OpenWhisk}}\right]}&= 2 \times \frac{E\left[\min\left(Z_1,Z_2\right)\right]}{E\left[ 
	\max\left( Z_1, Z_2 \right)\right]}\\
																																						&= \frac{1}{1.5} \\
																																						&\approx 0.67
	\end{split}
\end{equation*}
where $Z_1$ and $Z_2$ are i.i.d. exponentially distributed with mean $1$; the denominator follows from %
the fact that the expected response time of the workflow is equal to the delay of the slowest task in the pair of tasks, and %
the numerator follows from the fact that the delay until at least one task finishes is equal to the shortest delay in the %
pair. This almost perfect match to the theoretical performance benefits of %
replication in a high-availability deployment indicates that parallel node performance naturally becomes less correlated as systems %
increase in scale and we argue it suggests that Raptor will perform even better for workloads that have higher degrees of parallelism and in %
very large (warehouse) scale deployments.

\subsubsection{Word Count and Thumbnail Workflows}\label{ss:other-workloads}
\begin{figure}
	\centering
	\subfloat[Word Count\label{f:ch5-MRSteadyStateLatencies}]{
		\includegraphics[width=0.45\columnwidth]{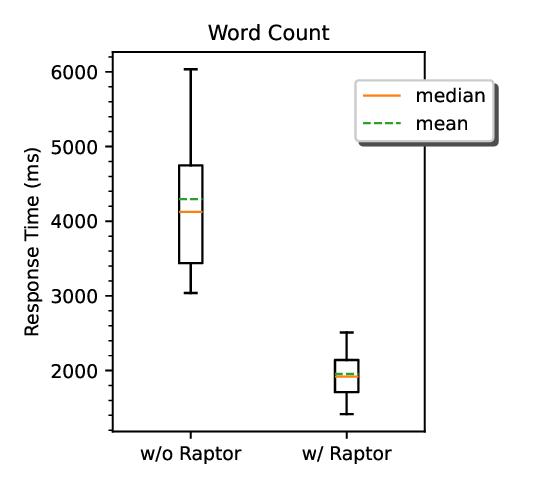}
	}
	\subfloat[Thumbnail Generation\label{f:ch5-WarmStarts}]{
		\includegraphics[width=0.45\columnwidth]{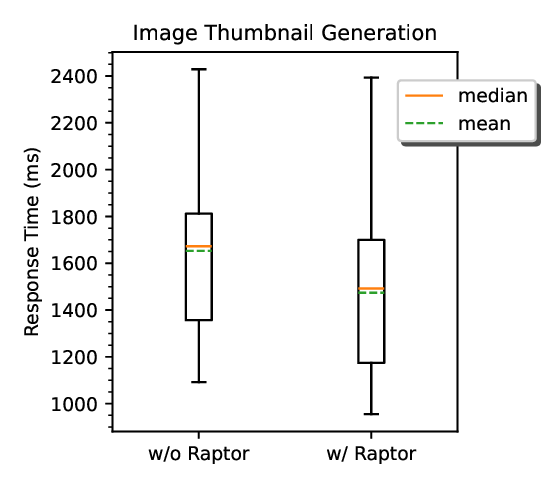}
	}
	\caption{Response Times for Other Workloads}
	\label{f:ch5-SteadyStateLatencies}
\end{figure}

Figure~\ref{f:ch5-SteadyStateLatencies} illustrates the performance of OpenWhisk with and without Raptor %
for some more realistic, but still synthetic, workloads. %
For thumbnail generation, we designed an action manifest similar %
to our SSH key generation manifest except with a concurrency of 4 instead of 2. The source image was first downloaded from a Google cloud %
storage bucket, then four thumbnails of different sizes were generated and subsequently uploaded back to the Google cloud storage %
bucket. For word count, we built an ad-hoc serverless map-reduce action sequence based on AWS's %
suggested architecture for ad-hoc data processing with serverless functions~\cite{liston_2016}. Raptor improves the performance %
of word count by over 50\% primarily because the distributed state-sharing stream can be used to send data for each phase of the %
workflow directly to reduce nodes rather than back through the OpenWhisk control datapath. While our thumbnail generation workflow %
demonstrates that for workloads that have more deterministic performance profiles, the benefits of Raptor over stock OpenWhisk %
are more muted but still positive. Both of these workflows were evaluated under a moderately loaded cluster deployed in a high-availability %
configuration.

\subsubsection{Reliability and Fault Tolerance}
\begin{figure}
	\centering
	\subfloat[w/ Raptor\label{f:ch5-JvTProbabilities}]{
		\includegraphics[width=0.45\columnwidth]{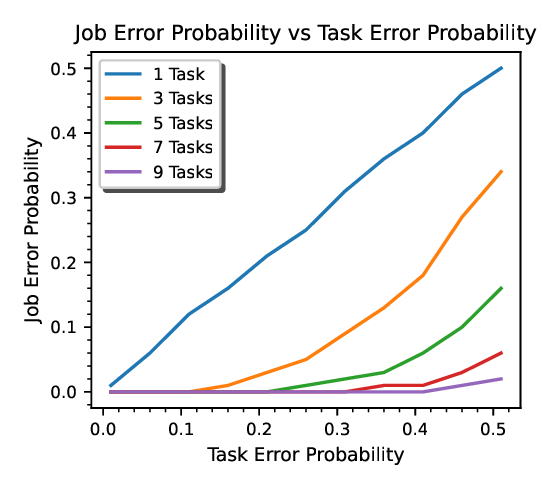}
	}
	\subfloat[w/o Raptor\label{f:ch5-JvTfjProbabilities}]{
		\includegraphics[width=0.45\columnwidth]{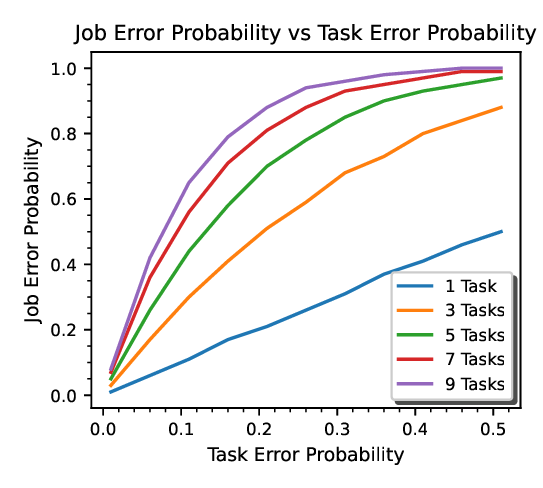}
	}
	\caption{Job vs Task Failure Probabilities }
	\label{f:ch5-reliability}
\end{figure}

As replication is a commonly used technique for improving application reliability; we illustrate in Figure~\ref{f:ch5-reliability} %
that, unlike the traditional fork-join approach, job failure probability versus parallel task failure probability will reduce %
for parallelizable jobs under Raptor as the number of parallel tasks increases. %
To generate these results, we simulate independent task failure probabilities on parallel nodes by designing an action manifest consisting %
of $N$ parallel busy waiting tasks that will either pass or fail at the end of a 100ms interval with a parameterized probability of error. %
Note that for stock OpenWhisk, the job failure probability for parallelizable jobs compounds with the number of parallel %
tasks as a typical \emph{fork-join} job requires that all $N$ parallel requests succeed. Raptor, however, only requires that %
\emph{at least one} of the $N$ parallel executors succeed, and so ensures that as the degree of parallelism in a job grows the %
job failure probability rapidly falls away to 0.

\section{Conclusions and Future Work}\label{s:conclusions}
Job scheduling at very large (warehouse) scales is a challenging problem, especially, for highly parallelizable %
data processing workflows like video processing~\cite{fouladi_encoding_2017}, linear algebra~\cite{shankar_serverless_2020}, or machine learning~\cite{wang_serverlessml_2019}. %
Traditional data processing frameworks like MapReduce~\cite{dean_mapreduce_2008}, Spark~\cite{zaharia_resilient_2012}, and Dryad~\cite{isard_dryad_2007} have a great deal of %
research directed at them which is focused on treating the unique and specific challenges of job and application scheduling at %
scale~\cite{dean_tail_2013, anan_effective_2013, ananthanarayanan_reining_2010}. Although serverless computing frameworks are different to data processing frameworks in the %
intended application; the control plane architecture of performance-oriented serverless computing frameworks like SAND~\cite{akkus_sand_2018} and %
Nightcore~\cite{jia_nightcore_2021} bears similarities to the architecture of data processing frameworks like MapReduce~\cite{dean_mapreduce_2008}, especially, in the use of %
distributed function executors. As such, we argue that scheduling techniques that are commonly adopted in these large-scale distributed data processing frameworks can also be %
adopted for serverless computing frameworks.

We present a distributed scheduling approach for serverless frameworks called Raptor that leverages the proven benefits of speculative %
execution~\cite{anan_effective_2013,ren_hopper_2015,anan_grass_2014} to greatly reduce the end-to-end response times of parallelizable serverless workflows. %
Unlike the executors used by other serverless frameworks~\cite{akkus_sand_2018, jia_nightcore_2021, carver_wukong_2020}, Raptor is designed to be able to %
take advantage of operating system job control mechanisms to facilitate much more granular scheduling of serverless function execution. For serverless workflows %
composed of more than one function, our distributed design uses a peer-to-peer communication channel to shorten the data pathway between subsequent user-defined %
functions in a workflow by directly passing function outputs between different instances of Raptor. For serverless workflows composed of parallelizable functions, Raptor %
uses this communication channel to share the current state-of-execution across all parallel instances of Raptor that are executing a particular workflow invocation. %
This both ensures that the functions in the workflow can be speculatively executed on multiple instances of Raptor, and that the excess resources consumed by %
speculatively execution are immediately freed as soon as \emph{any} of the parallel instances of execution produce the final outputs of the workflow rather than as soon as %
\emph{all} of the parallel instances of execution produce the final outputs of the workflow.

Our results demonstrate that for functions that have a certain degree of randomness built-in, like cryptographic key generation, Raptor nearly perfectly reproduces the theoretical %
performance gains that would be expected of \emph{mutually independent} function execution times but \emph{only} when the serverless framework has been %
deployed in a scaled-out and highly-available configuration. For smaller-scale deployments, at best, Raptor provides no performance benefits. %
We also demonstrate that for data-parallel workflows that are normally considered to have highly deterministic function execution times, %
like word counting and image processing, Raptor still provides non-negligible benefits for serverless frameworks deployed in such a highly-available configuration. Finally, we %
illustrate that in addition to \emph{reducing} job execution latencies, speculative execution techniques also \emph{reduce} job failure probabilities relative to %
the traditional fork-join approach.

\subsection{Future Work}
Raptor has been designed so that it can be used to \emph{extend} the scheduling capabilities of open source serverless computing frameworks like OpenWhisk, Knative, and %
OpenFaaS, with distributed function executors that can leverage traditional operating system job control mechanisms to schedule function execution at a much more granular level. %
However, due to the tight coupling between function execution and invocation queuing in Knative and OpenFaas, our prototype has only been integrated with OpenWhisk %
where a simple REST-based API is used to connect serverless function runtimes with the core OpenWhisk queuing and scheduling mechanisms. In its current %
iteration, our prototype of Raptor only supports POSIX job control signals and has only been compiled and tested on Linux, an implementation of Raptor that %
is more portable between different serverless frameworks and different operating systems needs to be developed to ensure that the benefits of distributed %
function executors can be realized by all serverless computing vendors. Furthermore, Raptor is designed to support the parallel execution of multiple user-defined functions %
by peering parallel instances of Raptor over the network to form distributed execution flights. Currently, our prototype only supports the SCTP transport protocol, however, %
a more robust solution should be able to support other transport protocols, like TCP, as well as additional application protocols such as HTTP/2 and WebSockets. %
Our prototype also does not currently support encryption using TLS, so the application state-sharing stream is unsecured, while this should not be a problem %
for an application that is designed to be deployed behind a firewall, a robust solution should support transport layer security to facilitate more flexible deployments of %
Raptor in serverless computing environments.

\bibliographystyle{plain}
\bibliography{IEEEabrv, references.bib}
\end{document}

%% file: figures/controller-arch.tex
\begin{tikzpicture}[
	rect/.style={
		rectangle,
		draw=black
	},
	queue/.style={
		cylinder,
		draw=black
	},
	point/.style={
		circle,
		inner sep=0pt,
		minimum size=0pt
	},
	external/.style={
		cloud,
		aspect=2,
		draw=black
	},
	hv path/.style={to path=-| (\tikztotarget)},
	vh path/.style={to path=|- (\tikztotarget)}
]
{
	\node(api)[rect, minimum height=60mm]{
		\begin{tabular}{c}
			Controller \\
			API
		\end{tabular}
	};
	\begin{scope}[local bounding box=AZ1]
		\node(c1)[rect, above right=of api.east, yshift=1em, xshift=1.5em]{Controller $1$};
		\node(s1)[rect, right=of c1, xshift=3em]{Scheduler $1$};
	\end{scope}
	\draw[black, dashed] ($(AZ1.south west) + (-1em, -1em)$) rectangle ($(AZ1.north east) + (1em, 1em)$);
	\node at ($(AZ1.north) + (0, 2em)$){Availability Zone 1};
	
	\node(d2)[below=of s1]{$\vdots$};
	\node(d1)[below=of c1]{$\vdots$};
	\node(p1)[point, minimum size=0.25em, fill=black, right=of d1, xshift=2.5em]{};	
	
	\begin{scope}[local bounding box=AZk]
		\node(ck)[rect, below=of d1]{Controller $k$};
		\node(sk)[rect, below=of d2]{Scheduler $k$};
	\end{scope}
	\draw[black, dashed] ($(AZk.south west) + (-1em, -1em)$) rectangle ($(AZk.north east) + (1em, 1em)$);
	\node at ($(AZk.south) + (0, -2em)$){Availability Zone $k$};
	
	\path[->, shorten <=0.5em, shorten >=0.5em] (api.east)  edge (c1.west)
															edge (ck.west);
															
	\path[->, shorten >=1em] (p1.base) edge (c1.east)
									   edge (ck.east)
									   edge (s1.west)
									   edge (sk.west);														
}
\end{tikzpicture}

%% file: figures/schedulerarch.tex
\begin{tikzpicture}[
	rect/.style={
		rectangle,
		draw=black
	},
	queue/.style={
		cylinder,
		draw=black
	},
	point/.style={
		circle,
		inner sep=0pt,
		minimum size=0pt
	},
	external/.style={
		cloud,
		aspect=2,
		draw=black
	},
	hv path/.style={to path=-| (\tikztotarget)},
	vh path/.style={to path=|- (\tikztotarget)}
]
{
	\begin{scope}[local bounding box=scheduler]
		\node(qm)[rectangle]{
			\begin{tabular}{|c|}\hline
				\scriptsize Queue 1 \\
				\hline
				$\vdots$ \\
				\hline
				\scriptsize Queue $n$ \\
				\hline
			\end{tabular}
		};
		\node[above=of qm, yshift=-2em]{Scheduler};
	\end{scope}
	\draw[black] ($(scheduler.north west) + (-0.5em, 0.5em)$) rectangle ($(scheduler.south east) + (0.5em, -0.5em)$);
	
	\node(controller)[rect, minimum height=12mm, left=of scheduler]{Controller};
	
	\node(i1)[rect, above right=of scheduler, yshift=-5em, xshift=2.5em]{Invoker $1$};
	\node[below=of i1, yshift=2.75em]{$\vdots$};
	\node(in)[rect, below right=of scheduler, yshift=5em, xshift=2.5em]{Invoker $M$};
	
	\node(p1)[point, minimum size=0.25em, fill=black, right=of scheduler]{};
	\path[->, shorten >=1em] (p1.base) edge (scheduler.east);
	\path[->, shorten >=0.5em] (p1.base) edge[vh path] (i1.west)
										 edge[vh path] (in.west);
	\path[<->, shorten >=0.5em, shorten <=1em] (scheduler.west) edge (controller.east);
}
\end{tikzpicture}

%% file: figures/InvokerPool.tex
\begin{tikzpicture}[
	rect/.style={
		rectangle,
		minimum size=12mm,
		draw=black
	},
	queue/.style={
		cylinder,
		minimum size=3mm,
		draw=black
	},
	point/.style={
		circle,
		inner sep=0pt,
		minimum size=0pt
	},
	external/.style={
		cloud,
		minimum size=6mm,
		aspect=2,
		draw=black
	},
	hv path/.style={to path=-| (\tikztotarget)},
	vh path/.style={to path=|- (\tikztotarget)}]
	\node(invoker)[rect]{Invoker};
	\node(io)[rect, left=of invoker]{Scheduler};
	\node(containers)[right=of invoker, label=north:Container Pool]{\begin{tabular}{|c|}
			\hline
			\scriptsize Container 1 \\
			\hline
			$\vdots$ \\
			\hline
			\scriptsize Container $N$ \\
			\hline		
	\end{tabular}};
	\node(k8s)[rect, below=of containers]{Container Orchestrator};
	
	\path[->, shorten <=0.5em, shorten >=0.5em] (invoker) edge[vh path] (k8s);
	\path[<->,shorten <=0.5em] (invoker) edge (containers);
	\path[->, shorten <=0.5em	] (k8s) edge (containers);
	\path[<->, shorten <=0.5em, shorten >=0.5em] (io) edge (invoker);
\end{tikzpicture}

%% file: figures/Raptor_Architecture.tex
\begin{tikzpicture}[
	rect/.style={
		rectangle,
		draw=black
	},
	vrect/.style={
		rectangle,
		draw=black,
		inner sep=0
	},
	hrect/.style={
		rectangle,
		draw=black,
		inner sep=0.5em
	},
	queue/.style={
		cylinder,
		draw=black
	},
	point/.style={
		circle,
		inner sep=0pt,
		minimum size=0pt
	},
	external/.style={
		cloud,
		aspect=2,
		draw=black
	},
	hv path/.style={to path=-| (\tikztotarget)},
	vh path/.style={to path=|- (\tikztotarget)},
	every edge quotes/.style={font=\footnotesize},
	ap label/.style={above right, inner sep=0.1em, pos=0.7},
	sapi label/.style={above, pos=0.8, inner sep=3em, to path=|- (\tikztotarget)}
	]
{	
	\def\bbinnersep{0.5em}
	\begin{scope}[local bounding box=flight]
		\node(s1)[rect]{\begin{tabular}{c}
			Raptor $0$ \\
			(Flight leader)
		\end{tabular}};
		\node(s3)[rect, right=of s1]{Raptor $1$};
		\node(p1)[point, right=of s3]{$\cdots$};
		\node(s4)[rect, right=of p1]{Raptor $N-2$};
		\node(s2)[rect, right=of s4]{Raptor $N-1$};
		\node(heading)[above=of p1, yshift=-3em]{Scheduling Flight};
	\end{scope}
	\draw[black, dashed] ($(flight.south west) + (-\bbinnersep, -\bbinnersep)$) rectangle ($(flight.north east) + (\bbinnersep, \bbinnersep)$);
	
	\def\vertspace{2em};
	\def\controlminwidth{6em};
	\def\controlminheight{12em};
	\def\labeloffset{5em};
	\def\ylabeloffset{6em};

	\begin{scope}[local bounding box=AZ2]
		\node(i2)[rect, minimum width=\controlminwidth, above=of p1, yshift=\vertspace]{Invoker $2$};
		\node(os2)[rect, minimum width=\controlminwidth, above=of i2]{Scheduler $2$};
		\node(c2)[rect, minimum width=\controlminwidth, above=of os2]{Controller $2$};
	\end{scope}
	\draw[black, dashed] ($(AZ2.south west) + (-\bbinnersep, -\bbinnersep)$) rectangle ($(AZ2.north east) + (\bbinnersep, \bbinnersep)$);
	\node[left=of AZ2, xshift=\labeloffset, yshift=\ylabeloffset]{\small{AZ 2}};

	\begin{scope}[local bounding box=AZ1]
		\node(i1)[rect, minimum width=\controlminwidth, left=of i2]{Invoker $1$};
		\node(os1)[rect, minimum width=\controlminwidth, above=of i1]{Scheduler $1$};
		\node(c1)[rect, minimum width=\controlminwidth, above=of os1]{Controller $1$};
	\end{scope}
	\draw[black, dashed] ($(AZ1.south west) + (-\bbinnersep, -\bbinnersep)$) rectangle ($(AZ1.north east) + (\bbinnersep, \bbinnersep)$);
	\node[left=of AZ1, xshift=\labeloffset, yshift=\ylabeloffset]{\small{AZ 1}};
	
	\begin{scope}[local bounding box=AZ3]
		\node(i3)[rect, minimum width=\controlminwidth, right=of i2]{Invoker $3$};
		\node(os3)[rect, minimum width=\controlminwidth, above=of i3]{Scheduler $3$};
		\node(c3)[rect, minimum width=\controlminwidth, above=of os3]{Controller $3$};
	\end{scope}
	\draw[black, dashed] ($(AZ3.south west) + (-\bbinnersep, -\bbinnersep)$) rectangle ($(AZ3.north east) + (\bbinnersep, \bbinnersep)$);
	\node[left=of AZ3, xshift=\labeloffset, yshift=\ylabeloffset]{\small{AZ 3}};
	
	\def\cpxshift{7em};
	\def\cpyshift{1em};
	\node(cp1)[vrect, minimum height=\controlminheight, left=of i1.west, xshift=-\cpxshift, yshift=\cpyshift]{\begin{tabular}{c}
		\small{Controller} \\
		\small{API}
	\end{tabular}};

	\node(cp2)[hrect, minimum width=2\controlminwidth, above=of c2]{Controller API};
	
	\def\sslen{0.5em};
	\def\lslen{2.5em};
	\def\bendangle{15};	
	\path[<->, bend right=\bendangle, shorten >=\sslen, shorten <=\sslen] (s1.south) edge (s3.south);
	\path[<->, bend right=\bendangle, shorten >=\sslen, shorten <=\sslen] (s1.south) edge (s4.south);
	\path[<->, bend right=\bendangle, shorten >=\sslen, shorten <=\sslen] (s1.south) edge["Flight Interconnects" below] (s2.south);
	\path[<->, bend right=\bendangle, shorten >=\sslen, shorten <=\sslen] (s3.south) edge (s4.south);
	\path[<->, bend right=\bendangle, shorten >=\sslen, shorten <=\sslen] (s3.south) edge (s2.south);
	\path[<->, bend right=\bendangle, shorten >=\sslen, shorten <=\sslen] (s4.south) edge (s2.south);
	
	\def\mslen{1.5em};
	\path[<->, shorten >=\mslen, shorten <=\sslen] (s1.north) edge (i1.south);
	\path[<->, shorten >=\mslen, shorten <=\sslen] (s3.north) edge (i2.south);
	\path[<->, shorten >=\mslen, shorten <=\sslen] (s4.north) edge (i2.south);
	\path[<->, shorten >=\mslen, shorten <=\sslen] (s2.north) edge["Action Interface" ap label] (i3.south);
	
	\def\rilen{0.75em};
	\path[<->] (i1.north) edge[shorten >=\rilen, shorten <=\sslen] (os1.south)
												edge[shorten >=\lslen, shorten <=\lslen] (os2.south);
	\path[<->] (i2.north) edge[shorten >=\lslen, shorten <=\lslen] (os1.south)
												edge[shorten >=\rilen, shorten <=\sslen] (os2.south)
												edge[shorten >=\lslen, shorten <=\lslen] (os3.south);
	\path[<->] (i3.north) edge[shorten >=\lslen, shorten <=\lslen] (os2.south)
												edge[shorten >=\rilen, shorten <=\sslen] (os3.south);
												
	\path[<->] (os1.north) 	edge[shorten >=\sslen, shorten <=\sslen] (c1.south)
													edge[shorten >=\lslen, shorten <=\lslen] (c2.south);
	\path[<->] (os2.north) 	edge[shorten >=\lslen, shorten <=\lslen] (c1.south)
													edge[shorten >=\sslen, shorten <=\sslen] (c2.south)
													edge[shorten >=\lslen, shorten <=\lslen] (c3.south);
	\path[<->] (os3.north) 	edge[shorten >=\lslen, shorten <=\lslen] (c2.south)
													edge[shorten >=\sslen, shorten <=\sslen] (c3.south);
	
	\path[<->, shorten >=\sslen, shorten <=\sslen, vh path] (cp1) edge (cp2);
	
	\def\capixoffset{8.9em}
	\def\lcapishort{0.75em}
	\path[<->, shorten >=\sslen, shorten <=\lcapishort] (c1.north) edge ($(cp2.south)+(-\capixoffset,0em)$);
	\path[<->, shorten >=\sslen, shorten <=\lcapishort] (c2.north) edge ($(cp2.south)+(0em,0em)$);
	\path[<->, shorten >=\sslen, shorten <=\lcapishort] (c3.north) edge ($(cp2.south)+(\capixoffset,0em)$);
	
	\def\capilabelxshift{2em}
	\draw[<->, shorten >=\sslen, shorten <=\sslen] (s1.north) |- node[above,xshift=-\capilabelxshift]{\footnotesize{API client interface}} ($(cp1.east) 
	+ (0,-3.4em)$);
	\path[<->, shorten >=\sslen, shorten <=\sslen, vh path] (s2.north) edge ($(cp1.east) + (0,-4.3em)$);
	\path[<->, shorten >=\sslen, shorten <=\sslen, vh path] (s3.north) edge ($(cp1.east) + (0,-3.7em)$);	
	\path[<->, shorten >=\sslen, shorten <=\sslen, vh path] (s4.north) edge ($(cp1.east) + (0,-4em)$);	
}
\end{tikzpicture}

%% file: figures/raptorInterfaces.tex
\begin{tikzpicture}[
	rect/.style={
		rectangle,
		draw=black
	},
	queue/.style={
		cylinder,
		draw=black
	},
	point/.style={
		circle,
		inner sep=0pt,
		minimum size=0pt
	},
	external/.style={
		cloud,
		aspect=4,
		draw=black,
		inner sep=0
	},
	hv path/.style={to path=-| (\tikztotarget)},
	vh path/.style={to path=|- (\tikztotarget)},
	]
{
	\begin{scope}[local bounding box=raptor]
		\node(heading)[point]{Raptor};
		\node(ap)[rect, below=of heading, yshift=3em]{Action Interface};
		\node(processes)[below=of ap]{\begin{tabular}{|l|}
			\hline
			\multicolumn{1}{|c|}{\textbf{User Functions}} \\
			\hline
			\texttt{fn1} \\
			\hline
			$\vdots$ \\
			\hline
			\texttt{fnN} \\
			\hline
		\end{tabular}};
	\end{scope}
	\def\padding{0.25em};
	\def\sval{0.5em};
	\draw[black] ($(raptor.south west) + (-\padding,-\padding)$) rectangle ($(raptor.north east) + (\padding, \padding)$);

	\def\owxwidth{13.5em};
	\def\apixshift{1em};
	\node(controller)[rect, minimum width=\owxwidth, left=of processes]{OpenWhisk Controller/API};
	\node(invoker)[rect, minimum width=\owxwidth, left=of ap, xshift=-\apixshift]{OpenWhisk Invoker};
	\node(core)[external, above left=of controller, yshift=-2.5em, dashed]{OpenWhisk Services};
	
	\def\rsval{1.4em};
	\def\lsval{0em};
	\path[<->, shorten <=0.2em] (ap) edge (processes);
	\path[<->, shorten <=\sval, shorten >=\rsval] (invoker) edge ($(ap.west) + (-0.5em,0)$);
	\path[<->, shorten <=\lsval, shorten >=\sval] ($(processes.west)+(-1em,0)$) edge (controller.east);
	\path[<->, bend left, dashed, shorten <=\sval, shorten >=\sval] (controller.west) edge (core.south);
	\path[<->, bend right, dashed, shorten <=\sval, shorten >=\sval] (invoker.west) edge (core.north);
}	
\end{tikzpicture}

%% file: figures/manifest-dag.tex
\begin{tikzpicture}
{
	\node(s){Start};
	\node[state, right=of s] (fn1){\texttt{fn1}};	
	\node[state, above right=of fn1] (fn2){\texttt{fn2}};
	\node[state, below right=of fn1] (fn3){\texttt{fn3}};
	
	\def\fxshift{3.5em};
	\node[state, right=of fn1, xshift=\fxshift] (fn4) {\texttt{fn4}};
	\node[right=of fn4](f){Finish};
	
	\def\sv{0.5em}
	\def\bangle{10}
	\path[->, bend left=\bangle, shorten >=\sv, shorten <=\sv] (s) 		edge (fn1);
	\path[->, bend left=\bangle, shorten >=\sv, shorten <=\sv] (fn1) 	edge (fn2)
																																		edge (fn3);
	\path[->, bend left=\bangle, shorten >=\sv, shorten <=\sv] (fn2) 	edge (fn4);
	\path[->, bend left=\bangle, shorten >=\sv, shorten <=\sv] (fn3) 	edge (fn4);
	\path[->, bend left=\bangle, shorten >=\sv, shorten <=\sv] (fn4) 	edge (f);
	
	\path[<-, red, bend right=\bangle, shorten >=\sv, shorten <=\sv] (s) 		edge (fn1);
	\path[<-, red, bend right=\bangle, shorten >=\sv, shorten <=\sv] (fn1) 	edge (fn2)
																																					edge (fn3);
	\path[<-, red, bend right=\bangle, shorten >=\sv, shorten <=\sv] (fn2) 	edge (fn4);
	\path[<-, red, bend right=\bangle, shorten >=\sv, shorten <=\sv] (fn3) 	edge (fn4);
	\path[<-, red, bend right=\bangle, shorten >=\sv, shorten <=\sv] (fn4) 	edge (f);
	
	\def\lyshift{3em};
	\def\llyshift{1.5em};
	\node[below=of f, yshift=\lyshift](p){};
	\node[below left=of p](l1){};
	\node[right=of l1](f1){Execution Direction};
	\node[below=of l1, yshift=\llyshift](l2){};
	\node[right=of l2](f2){Search Direction};
	
	\path[->] 			(l1) edge (f1);
	\path[->, red] 	(l2) edge (f2);
}
\end{tikzpicture}